\documentclass[twocolumn]{svjour3}

\usepackage{cite}
\usepackage{amsmath}
\usepackage{graphicx}

\begin{document}

\title{Scalable Quantum Computing Architecture with Mixed Species Ion Chains}
\author{John Wright \and Carolyn Auchter \and Chen-Kuan Chou \and Richard D. Graham \and Thomas W. Noel \and Tomasz Sakrejda \and Zichao Zhou \and Boris B. Blinov}
\institute{ 
	John Wright \and Carolyn Auchter \and Chen-Kuan Chou \and Richard D. Graham \and Thomas W. Noel \and Tomasz Sakrejda \and Zichao Zhou \and Boris B. Blinov \at Department of Physics, University of Washington, Seattle, WA 98195 \\
	\and John Wright \at \email{johnwri@uw.edu}
}
\date{\today}
\journalname{Quantum Information Processing}

\maketitle

\begin{abstract}
	We report on progress towards implementing mixed ion species quantum information processing for a scalable ion trap architecture.  Mixed species chains may help solve several problems with scaling ion trap quantum computation to large numbers of qubits.  Initial temperature measurements of linear Coulomb crystals containing barium and ytterbium ions indicate that the mass difference does not significantly impede cooling at low ion numbers.  Average motional occupation numbers are estimated to be \(\bar{n} \approx 130\) quanta per mode for chains with small numbers of ions, which is within a factor of three of the Doppler limit for barium ions in our trap.  We also discuss generation of ion-photon entanglement with barium ions with a fidelity of \(F \ge 0.84\), which is an initial step towards remote ion-ion coupling in a more scalable quantum information architecture.  Further, we are working to implement these techniques in surface traps in order to exercise greater control over ion chain ordering and positioning. 
\keywords{Ion Trapping \and Sympathetic Cooling \and Mixed Species Ion Chains \and Scalable Quantum Computing Architecture}
\end{abstract}
	
\section{Introduction}
Trapped ions are a promising technology for quantum computing and many demonstrations of the necessary techniques have already been made including long coherence times\cite{Olmschenk:07}, fast readout\cite{Olmschenk:07}, and entangling ion-ion gates\cite{Kirchmair:09, Hayes:10}.  Useful quantum information processing with trapped ions requires scaling these techniques to large numbers of ions.  One possible method for addressing the difficulties in this scaling is the Modular Universal Scalable Ion-Trap Quantum Computer (MUSIQC) architecture\cite{Monroe:14}, where multiple ion traps containing tens of ions are coupled together via photon-mediated ion-ion entanglement.  We are working on many fronts to implement this architecture using mixed species ion chains of  \(^{138}\)Ba\(^+\) and \(^{171}\)Yb\(^+\) to address problems with field crosstalk and ion temperature.

	In the MUSIQC architecture, an expandable number of Elementary Logic Units (ELUs), microfabricated traps holding linear chains of 10 to 100 ions, are linked together using a photonic interface to form a large-scale system\cite{Monroe:14}. Local quantum gates are performed using motional coupling between ions in the same trap. Several ions in each chain are reserved for performing a slower entanglement operation between ions in different ion traps coupled by optical fibers.  This long distance entanglement will be accomplished using photon-mediated ion-ion entanglement, in which a pair of ions are projected into an entangled Bell state by a combined measurement of their emitted single photons\cite{Moehring:07}.  The necessary photon measurements can be performed using fiber switches and fiber beamsplitters once both traps are coupled to optical fibers.  We intend to separate the fast motional coupling and the slower remote ion entanglement to different ion species whose atomic transitions are widely separated in frequency.  Attempting remote entanglement generation and laser cooling ions of one species then will not cause decoherence in ions of the other species.

	Quantum information can be stored in the ground state Zeeman levels of \(^{138}\)Ba\(^+\) ions and the ground-state, magnetic-field-insensitive hyperfine levels of \linebreak\(^{171}\mathrm{Yb}^+\).  The hyperfine levels of ytterbium are insensitive to most environmental noise and coherence times of several seconds can be easily achieved\cite{Olmschenk:07}.  The magnetically sensitive Zeeman levels of barium are poor qubits for long term quantum information storage, but the qubit state is correlated with the polarization and frequency of the 493~nm emitted photons.  Therefore, the photon-mediated remote entanglement operation can easily be performed between qubits of this type.  Once entanglement is created between remote barium ions, it can be transferred to the stable ytterbium hyperfine levels using motional gates and then used during  quantum computation with the ytterbium ions.  Since quantum information is not stored in barium ions for an extended period of time, they can be laser cooled without causing decoherence of the entire system.  Using barium for this task is advantageous because its relevant atomic transition is at a relatively long wavelength, 493~nm, among ion species that can be easily trapped and laser cooled, which allows for better fiber transmission efficiencies.  Either a polarization or a frequency photonic qubit is possible, depending on other experimental concerns\cite{Luo:09}.

	Readout and initialization of the qubit state is easily performed in \(^{171}\)Yb\(^+\) because of its hyperfine structure\cite{Olmschenk:07}.  Either task can be accomplished by switching the microwave drive of the Electro-Optic Modulator (EOM) that provides the frequency sideband that addresses the F=0 hyperfine state (see Figure 1).  When the frequency modulation is turned off, ions will be optically pumped into the F=0, m\(_F\)=0 qubit state.  State detection can also be implemented straightforwardly using this microwave switch by detecting 369~nm fluorescence emitted while the EOM is disabled and the F=0 level is outside the cooling cycle.  

\section{Experimental Setup}
We simultaneously trap \(^{138}\)Ba\(^+\) and \(^{174}\)Yb\(^+\) ions in a 4-rod linear Paul trap.  Although \(^{171}\)Yb\(^+\) will be used in quantum information experiments we are currently working with \(^{174}\)Yb\(^+\) due to experimental limitations.  The mass difference is close enough that the effects on the normal mode structure are similar in both isotopes.  The barium ions are laser cooled and detected using a frequency-doubled 986~nm External Cavity Diode Laser (ECDL) and a 650~nm ECDL (see Figure 1).  Ytterbium ions are then loaded and are not addressed by any lasers.  They are cooled sympathetically via Coulomb interactions with the Doppler-cooled barium ions.  An Electron Multiplying CCD (EMCCD) camera  images the 493~nm fluorescence of the barium ions, while the positions of the ytterbium ions can be inferred from the location of gaps in the barium ion chain.  The species of the dark ions has been confirmed to be ytterbium by measuring the mass shift of the secular frequencies.  Figure 2 shows an example of images collected from such a mixed-species ion chain.

\begin{figure}[ht!]
	\centering
	\includegraphics[width=0.38\textwidth]{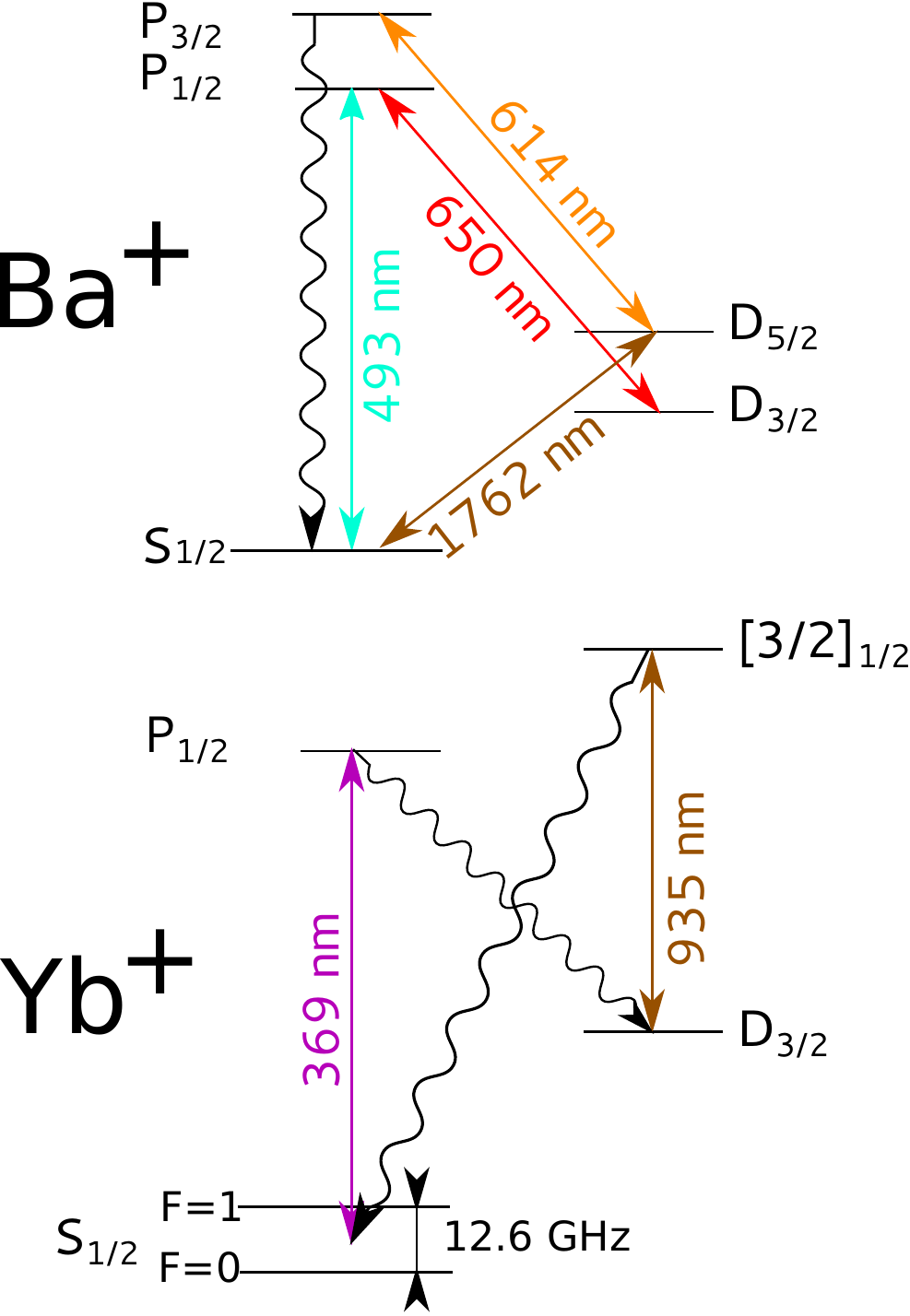}
	\caption{Relevant energy level structure of barium and ytterbium (not to scale).  Barium is Doppler cooled by 493~nm and 650~nm light, while state detection is accomplished by ``shelving'' using a 1762~nm source and ``deshelving'' using a 614~nm source.  The ytterbium hyperfine levels are only present in \(^{171}\)Yb\(^+\) and not \(^{174}\)Yb\(^+\) because of their nuclear spins.  These levels can be initialized and detected using 369~nm and 935~nm light.}
\end{figure}

\begin{figure}[ht!]
	\centering
	\includegraphics[width=0.48\textwidth]{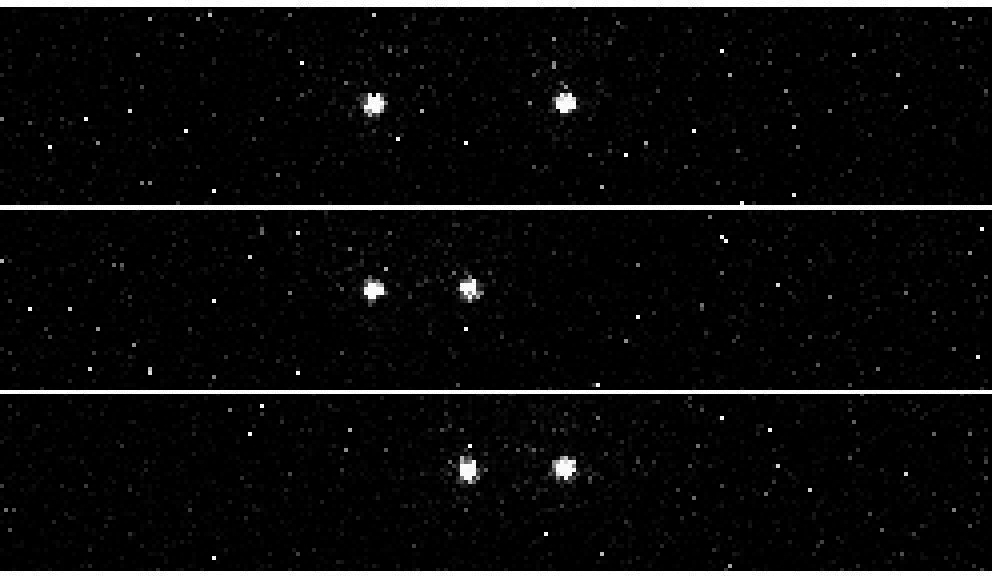}
	\caption{Typical images from a chain of two \(^{138}\)Ba\(^+\) ions and one \(^{174}\)Yb\(^+\) ion.  The barium ions are fluorescing under laser cooling, while the ytterbium ion is dark.  The Yb\(^+\) ion is in the middle in the top panel, the right in the middle panel, and the left in the bottom panel.}
\end{figure}
	
	In barium, the initial ground Zeeman state is prepared by polarization instead of frequency-based optical pumping.  For example, \(\sigma^+\) polarized 493~nm light will drive ground state population in barium ions to the m\(_J\)=\(+\frac{1}{2}\) Zeeman qubit state.  We use a simple polarization technique to reduce laser power requirements.  Only one 493~nm beam is incident on the trapped ions and its polarization can rapidly be switched from linear to circular by a Pockels cell to perform either cooling or initialization, respectively.  State detection of the barium Zeeman states is accomplished by selectively transferring or ``shelving'' one ground Zeeman state to the \(5D_{5/2}\) level using a narrow 1762~nm laser (see Figure 1).  The \(5D_{5/2}\) level of barium is outside of the cooling cycle and has a long lifetime of approximately 30 seconds\cite{Gurell:07}, which allows easy detection of successful shelving events by collecting fluorescence. Single qubit operations can be performed on either species by applying coherent rf at appropriate frequencies (10-100~MHz for \(^{138}\)Ba\(^+\), depending on the applied magnetic field, and about 12.6~GHz for \(^{171}\)Yb\(^+\)) or by driving stimulated Raman transitions.

	A modelocked 1064~nm Nd:YVO4 laser will be used to perform stimulated Raman transitions and transfer remote ion-ion entanglement from barium to ytterbium.  This laser will also be used to perform entangling gates between multiple ytterbium ions as a part of quantum computing protocols.  The bandwidth of the pulses (approximately 100~GHz) can easily span the 12.6~GHz hyperfine splitting of ytterbium to drive Raman transitions between these states.  In order to efficiently address both barium and ytterbium, the second (532~nm) and third (355~nm) harmonics of the laser will be used.  The third harmonic is ideal for minimizing photon scattering and Stark shifts in ytterbium operations\cite{Campbell:10}, and the second harmonic is expected to be near enough to barium's transitions to efficiently drive Raman operations.  The modelocked laser was home-built and produces 2~W of 1064~nm light.  It is modelocked using a semiconductor saturable absorber based on \cite{Schlatter:04, Sun:10} and produces 17~ps pulses, as measured by an autocorrelator, at a repetition rate of 150~MHz.  Its output will be power and repetition rate stabilized by feedback to two acousto-optic modulators used to create the frequency separation necessary for Raman transitions\cite{Hayes:10}.  

\section{Results}
\subsection{Mixed Species Chains}
	To utilize the advantages of this mixed species quantum computing proposal it is necessary to maintain a low ion chain temperature while cooling only one species of ion.  We have begun investigating both the fraction and ordering of laser cooled ions necessary to maintain this condition.  The ordering of the species in our trap is currently random, but could be controlled in the future through the use of a microfabricated surface trap with many DC control electrodes \cite{Shu:14, Wright:13}.  

\begin{figure}
	\centering
	\input{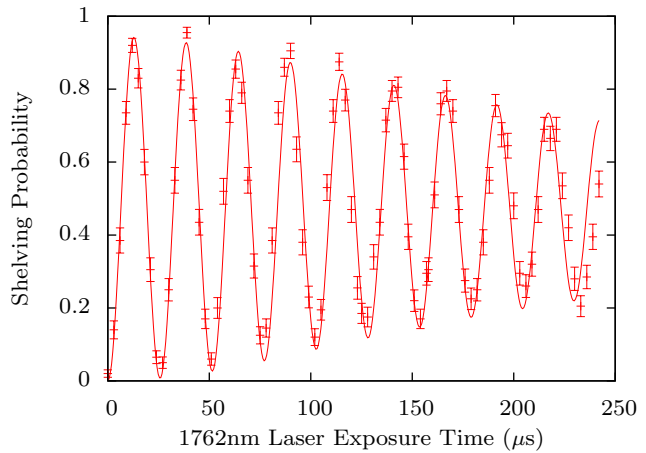}
	\caption{Carrier Rabi flops on the 1762~nm transition in a single barium ion. The Rabi flops fit to a total radial motional occupation number of \(\Sigma_i \bar{n}_i\) = \(\bar{n}_x + \bar{n}_y\) = 129(11)~quanta.}
\end{figure}

\begin{figure}
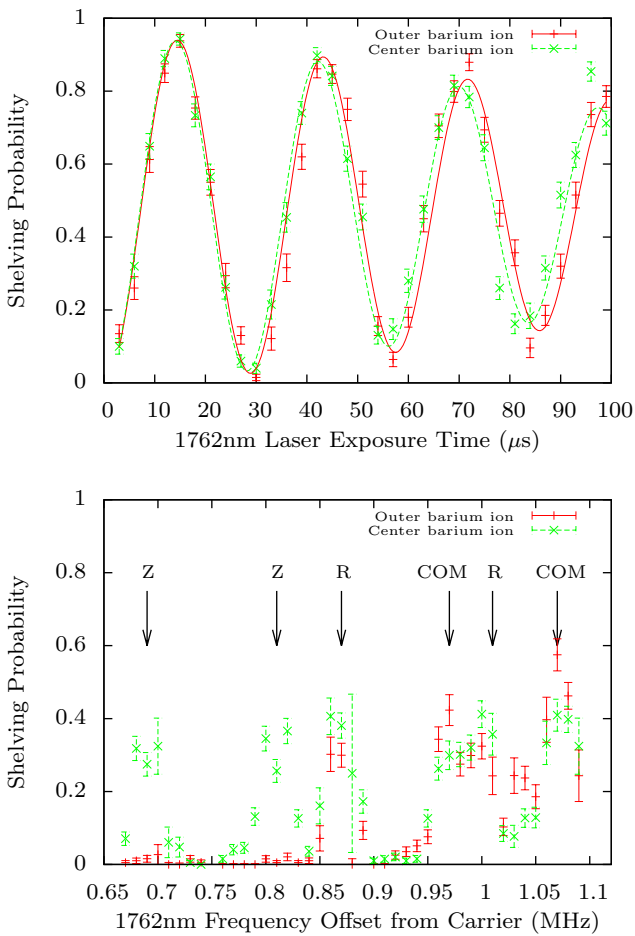

	\centering
	\input{Fig4a.tex}
	\input{Fig4b.tex}
	\caption{Carrier Rabi flops (top) and a scan over the radial mode frequencies (bottom) on the 1762~nm transition of barium in an ion chain with two barium ions and a single ytterbium ion. The Rabi flop decay fits to a total radial motional occupation number of 247(34)~quanta and 280(37)~quanta for the outer and center barium ion, respectively.  The radial mode scan shows the expected 6 modes per ion and their relative strengths are within a factor of two from the predicted eigenvector components.}
\end{figure}

	In chains with a large number of ions, it is difficult and time consuming to measure the thermal occupation number of each mode through direct sideband measurements.  Instead, we have been measuring the approximate temperature of the entire chain through two different types of measurements.  The first method relies on the decay of Rabi flop contrast of a carrier transition.  At the temperature these chains currently reach, this decay is mainly due to driving many motional occupation number carrier transitions with different Rabi frequencies rather than laser noise.  As a result, the shelving probability (\(P_\mathrm{shelved}\)), assuming a single addressed motional degree of freedom (DOF) and that we are in the Lamb-Dicke limit, can be fit to the expression:
\begin{eqnarray}
P_\mathrm{shelved} & = & \frac{1}{2}\left( 1 - \sum\limits_{n=0}^\infty p_n \cos( 2 \Omega_{n,n} t ) \right)\mathrm{, with} \\
p_{n} & = & \frac{1}{\bar{n}+1}\left(\frac{\bar{n}}{\bar{n}+1}\right)^n \\
\Omega_{n,n} & = & \Omega_0 L_n( \eta^2 ) \approx \Omega_0 (1 - \eta^2 n),
\end{eqnarray}
where \(p_n\) is the probability of occupation of the \(n^\mathrm{th}\) excited motional level, \(\bar{n}\) is the average motional occupation number, \(\Omega_{n,n}\) is the Rabi frequency for the carrier transition from motional excited level \(n\), \(\Omega_0\) is the Rabi frequency for the ground motional state carrier transition, \(L_n\) is the Laguerre polynomial of order \(n\), and \(\eta\) is the Lamb-Dicke parameter for the motional mode.  This expression can be extended to multiple motional DOF, and a parameter \(\Sigma_i \eta_i^2 \bar{n}_i\), where the sum is over addressed motional modes of the chain, can be fit from the data.  Due to the 1762~nm beam alignment in our experiment, only the radial modes contribute, and there are 2 DOF per ion in the chain.  If all the Lamb-Dicke parameters for the radial modes are similar, we can approximate them as equal in order to fit \(\Sigma_i \bar{n}_i\) to the data.  

	In our trap a single barium ion has \(\Sigma_i \bar{n}_i\) = \(\bar{n}_x + \bar{n}_y\) = 129(11)~quanta, corresponding to a temperature of 3.4~mK in our 1.1~MHz radial trap (see Figure 3). We have also collected data using a chain of two barium ions and one ytterbium ion.  The radial mode structure for this chain is different depending on the configuration of the different species, so we have only collected data while the ytterbium is one of the outer ions.  We measure higher minimum average occupation numbers of 247(34)~quanta for the outer barium ion and 280(37)~quanta for the center barium ion.  In our analysis we have again assumed that the radial modes have the same frequency even though the measured radial frequencies differ by more than 30 percent.  Therefore these occupation numbers should be understood as approximations to the actual temperature for rough comparison purposes.  Based on the expected eigenvector components for the modes, if the quanta are evenly distributed between the modes the average occupation number is \(\bar{n} \approx 130\) per mode.

	Analyzing the normal modes of these chains using classical mechanics\cite{Home:11}, we expect some radial modes to couple more strongly to one species than the other.  When modes couple strongly to ytterbium and weakly to barium it becomes more difficult to cool them using barium cooling lasers, and we expect this to be an ongoing problem as we increase the ion number in our trap.  For a chain of two barium and one ytterbium, we predict that the four higher frequency radial modes, the common (COM) and rocking (R) modes, should couple strongly to both barium ions, but the two lower frequency modes, the zigzag (Z) modes, should couple strongly to just the ytterbium ion.  In the bottom panel of figure 4, the Z mode peaks at 0.69~MHz and 0.81~MHz are very small for the outer barium ion, meaning that this ion barely couples to these modes.  The center barium ion has approximately the same transition strength for all six modes.  Because the eigenvector component for the center barium ion to participate in the Z modes is actually a factor of two smaller than the components for the COM and R modes, we believe the lower frequency mode has a significantly higher occupation number. Therefore the average \(\bar{n}\) given above is probably not evenly distributed between the modes and the COM and R modes probably have average occupation numbers more similar to the case of a single barium ion.  Finding techniques to efficiently cool these decoupled modes will be an ongoing challenge in working with this type of system.

\begin{figure}
	\centering
	\input{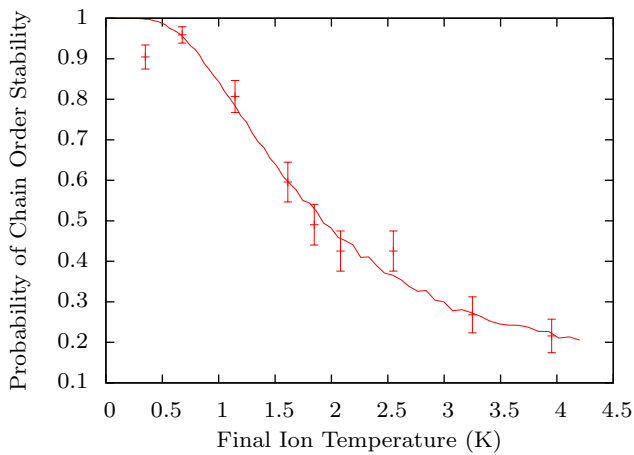}
	\caption{Probability of a chain of seven barium ions and one ytterbium ion not detectably reordering is plotted as a function of final chain temperature.  The temperature is found by fitting the data taken as a function of dark time to the simulated curve (shown by the solid line) assuming a linear heating rate.  The initial temperature is found to be 206~mK and the heating rate is 0.46~K/s.}
\end{figure}

	An alternative method for approximating the temperature and heating rate of the ion chains involves measuring the probability of reordering of the barium and ytterbium ions in the chain as a function of uncooled dark time during which the ions heat due to electric field noise.  Numerically simulating the classical, differential equation of motion for this probability over random initial conditions with a given temperature shows that the probability has a steep slope as a function of ion temperature.  Collecting data on this slope functions as a diagnostic temperature measurement that can be used to fine tune experimental parameters.  This measurement is much faster than collecting data for Rabi flops and fitting to find the temperature from them.  Figure 5 shows the collected data and the simulation, where the temperature of the data points has been found by performing a least squares fit to the simulation using an initial temperature and linear heating rate as free parameters.  The initial temperature fits to 206~mK and the heating rate fits to 0.46~K/s. This data was collected before our cooling was substantially improved to collect Rabi flops and shows a very high temperature.

\subsection{Planar Surface Ion Traps}
	Implementing this system using planar surface traps (including those designed and produced by GTRI\cite{Wright:13}, Sandia National Laboratory\cite{Allcock:12}, and others\cite{Daniilidis:11}), has several advantages.  The large number of independent DC control voltages allows for the fine positioning of ions, and allows for the shuttling of ions between different trapping regions.  We have successfully demonstrated loading chains of barium ions in a separate loading region while also cooling ions in a ``quantum'' information region in the Sandia Y-trap.  The loading region can serve as a reserve to replace ions lost through background gas collisions.  Arbitrarily ordered chains of barium and ytterbium can be built up from separate loading or storage regions for barium and ytterbium.  We have implemented a simple technique for controlling the necessary DC electrodes using a FPGA-controlled digital to analog converter (DAC) system using DACs with a large number of output channels (32) to reduce cost\cite{Graham:14}.

\subsection{Ion-Photon Entanglement}
	The first step toward performing remote ion-ion entanglement is to demonstrate collection and measurement of photons entangled with the state of a trapped ion.  We have recently demonstrated creation of a maximally entangled ion-photon Bell state with fidelity \(F \ge 0.84(1)\) \cite{Auchter:14}.  The probability, \(P\), for achieving and measuring this entanglement during each experimental run is given by:
\begin{equation}
	P = P_\mathrm{exc} f  \eta \frac{\Omega}{4 \pi} f_\mathrm{gate} T
\end{equation}
where \(P_\mathrm{exc}\) is the probability of excitation (\(\approx\) 0.2 in our experiment), \(f\) is the branching ratio to ground state from the \(6P_{1/2}\) state of Ba\(^+\) (\(\approx\) 0.75), \(\eta\) is the quantum efficiency of photon detection (\(\approx\) 0.2), \(\Omega\) is the fluorescence collection solid angle (\(\approx 0.02 * 4 \pi\)), \(f_\mathrm{gate}\) is the fraction of emitted photons in our detection gating (\(\approx\) 0.8), and \(T\) is the transmission through imaging optics (\(\approx\) 0.3).  These limit the ion-photon entanglement rate to \(\approx\) 2.5~Hz given our repetition rate of approximately 17~kHz.  However,  the resulting entanglement rate for remote ion-ion entanglement will be much lower because two ions must be simultaneously entangled with photons, and therefore these limiting factors are squared.  The easiest factor to improve in this expression is the collection solid angle through the use of optical cavities in vacuum \cite{Sterk:12}, diffractive optics \cite{Clark:14, Jechow:11}, or improved collection optics.   We have designed an ion trap incorporating a parabolic mirror similar to our previous spherical mirror design\cite{Shu:11}.  Using this trap, we have measured light collection efficiencies as high as \(\Omega = 0.4 * 4 \pi\), which will improve the remote entanglement rate by a factor of 400.  Further, we can improve the excitation probability to unity by using a frequency doubled modelocked Ti:Sapphire laser.  We have separately coupled these entangled photons to remote traps using commercial optical fiber (Nufern HP-460) and have demonstrated that the necessary interference measurement to generate ion-ion entanglement can be performed using commercial fiber beam splitters (Thorlabs FC488-50B-APC).  

\section{Conclusions}
Barium and ytterbium both have useful properties for realizing different parts of the MUSIQC quantum computing architecture.  Using both species simultaneously allows tasks to be separated between the two species to avoid problems with field crosstalk and ion heating.  We have successfully trapped both species simultaneously and implemented state detection, initial temperature measurements and barium ion-photon entanglement.   We are working towards implementing entangling gates between the different ion species in a single trap as well as between remote barium ions via photon-mediated ion-ion entanglement.

\begin{acknowledgements}
The authors would like to thank Matthew R. Hoffman, Spencer R. Williams, and Anupriya Jayakumar for useful conversations.  We would also like to acknowledge support from the Intelligence Advanced Research Projects Activity through the Multi-Qubit Coherent Operations Program and the National Science Foundation under grant number PHY-1067054.
\end{acknowledgements}

\bibliographystyle{unsrt}
\bibliography{paper}{}

\begin{thebibliography}{10}

\bibitem{Olmschenk:07}
S.~Olmschenk, K.~C. Younge, D.~L. Moehring, D.~N. Matsukevich, P.~Maunz, and
  C.~Monroe.
\newblock Manipulation and detection of a trapped {Y}b$^{+}$ hyperfine qubit.
\newblock {\em Phys. Rev. A}, 76:052314, Nov 2007.

\bibitem{Kirchmair:09}
G~Kirchmair, J~Benhelm, F~Zähringer, R~Gerritsma, C~F Roos, and R~Blatt.
\newblock Deterministic entanglement of ions in thermal states of motion.
\newblock {\em New Journal of Physics}, 11(2):023002, 2009.

\bibitem{Hayes:10}
D.~Hayes, D.~N. Matsukevich, P.~Maunz, D.~Hucul, Q.~Quraishi, S.~Olmschenk,
  W.~Campbell, J.~Mizrahi, C.~Senko, and C.~Monroe.
\newblock Entanglement of atomic qubits using an optical frequency comb.
\newblock {\em Phys. Rev. Lett.}, 104:140501, Apr 2010.

\bibitem{Monroe:14}
C.~Monroe, R.~Raussendorf, A.~Ruthven, K.~R. Brown, P.~Maunz, L.-M. Duan, and
  J.~Kim.
\newblock Large-scale modular quantum-computer architecture with atomic memory
  and photonic interconnects.
\newblock {\em Phys. Rev. A}, 89:022317, Feb 2014.

\bibitem{Moehring:07}
D.~L. Moehring, P.~Maunz, S.~Olmschenk, K.~C. Younge, D.~N. Matsukevich, L.-M.
  Duan, and C.~Monroe.
\newblock Entanglement of single-atom quantum bits at a distance.
\newblock {\em Nature}, 449(7158):68--71, Sep 2007.

\bibitem{Luo:09}
L.~Luo, D.~Hayes, T.A. Manning, D.N. Matsukevich, P.~Maunz, S.~Olmschenk, J.D.
  Sterk, and C.~Monroe.
\newblock Protocols and techniques for a scalable atom-photon quantum network.
\newblock {\em Fortschritte der Physik}, 57(11-12):1133--1152, 2009.

\bibitem{Gurell:07}
J.~Gurell, E.~Bi\'emont, K.~Blagoev, V.~Fivet, P.~Lundin, S.~Mannervik, L.-O.
  Norlin, P.~Quinet, D.~Rostohar, P.~Royen, and P.~Schef.
\newblock Laser-probing measurements and calculations of lifetimes of the
  $5d\phantom{\rule{0.2em}{0ex}}^{2}d_{3∕2}$ and
  $5d\phantom{\rule{0.2em}{0ex}}^{2}d_{5∕2}$ metastable levels in
  $\mathrm{Ba}\phantom{\rule{0.2em}{0ex}}ii$.
\newblock {\em Phys. Rev. A}, 75:052506, May 2007.

\bibitem{Campbell:10}
W.~C. Campbell, J.~Mizrahi, Q.~Quraishi, C.~Senko, D.~Hayes, D.~Hucul, D.~N.
  Matsukevich, P.~Maunz, and C.~Monroe.
\newblock Ultrafast gates for single atomic qubits.
\newblock {\em Phys. Rev. Lett.}, 105:090502, Aug 2010.

\bibitem{Schlatter:04}
Adrian Schlatter, S.~C. Zeller, R.~Grange, R.~Paschotta, and U.~Keller.
\newblock Pulse-energy dynamics of passively mode-locked solid-state lasers
  above the q-switching threshold.
\newblock {\em J. Opt. Soc. Am. B}, 21(8):1469--1478, Aug 2004.

\bibitem{Sun:10}
L~Sun, L~Zhang, H~J Yu, L~Guo, J~L Ma, J~Zhang, W~Hou, X~C Lin, and J~M Li.
\newblock 880 nm ld pumped passive mode-locked {TEM} 00 {Nd:YVO} 4 laser based
  on sesam.
\newblock {\em Laser Physics Letters}, 7(10):711, 2010.

\bibitem{Shu:14}
G.~Shu, G.~Vittorini, A.~Buikema, C.~S. Nichols, C.~Volin, D.~Stick, and
  Kenneth~R. Brown.
\newblock Heating rates and ion-motion control in a $\mathsf{Y}$-junction
  surface-electrode trap.
\newblock {\em Phys. Rev. A}, 89:062308, Jun 2014.

\bibitem{Wright:13}
Kenneth Wright, Jason~M Amini, Daniel~L Faircloth, Curtis Volin, S~Charles
  Doret, Harley Hayden, C-S Pai, David~W Landgren, Douglas Denison, Tyler
  Killian, Richart~E Slusher, and Alexa~W Harter.
\newblock Reliable transport through a microfabricated {X}-junction
  surface-electrode ion trap.
\newblock {\em New Journal of Physics}, 15(3):033004, 2013.

\bibitem{Home:11}
J~P Home, D~Hanneke, J~D Jost, D~Leibfried, and D~J Wineland.
\newblock Normal modes of trapped ions in the presence of anharmonic trap
  potentials.
\newblock {\em New Journal of Physics}, 13(7):073026, 2011.

\bibitem{Allcock:12}
D.T.C. Allcock, T.P. Harty, H.A. Janacek, N.M. Linke, C.J. Ballance, A.M.
  Steane, D.M. Lucas, Jr. Jarecki, R.L., S.D. Habermehl, M.G. Blain, D.~Stick,
  and D.L. Moehring.
\newblock Heating rate and electrode charging measurements in a scalable,
  microfabricated, surface-electrode ion trap.
\newblock {\em Applied Physics B}, 107(4):913--919, 2012.

\bibitem{Daniilidis:11}
N~Daniilidis, S~Narayanan, S~A Möller, R~Clark, T~E Lee, P~J Leek, A~Wallraff,
  St~Schulz, F~Schmidt-Kaler, and H~Häffner.
\newblock Fabrication and heating rate study of microscopic surface electrode
  ion traps.
\newblock {\em New Journal of Physics}, 13(1):013032, 2011.

\bibitem{Graham:14}
R.~D. Graham, S.-P. Chen, T.~Sakrejda, J.~Wright, Z.~Zhou, and B.~B. Blinov.
\newblock A system for trapping barium ions in a microfabricated surface trap.
\newblock {\em AIP Advances}, 4(5), 2014.

\bibitem{Auchter:14}
Carolyn Auchter, Chen-Kuan Chou, Thomas~W. Noel, and Boris~B. Blinov.
\newblock Ion-photon entanglement and bell inequality violation with
  138{B}a$+$.
\newblock {\em J. Opt. Soc. Am. B}, 31(7):1568--1572, Jul 2014.

\bibitem{Sterk:12}
J.~D. Sterk, L.~Luo, T.~A. Manning, P.~Maunz, and C.~Monroe.
\newblock Photon collection from a trapped ion-cavity system.
\newblock {\em Phys. Rev. A}, 85:062308, Jun 2012.

\bibitem{Clark:14}
Craig~R. Clark, Chin-wen Chou, R.~Ellis, A.\, Jeff Hunker, Shanalyn~A. Kemme,
  Peter Maunz, Boyan Tabakov, Chris Tigges, and Daniel~L. Stick.
\newblock Characterization of fluorescence collection optics integrated with a
  microfabricated surface electrode ion trap.
\newblock {\em Phys. Rev. Applied}, 1:024004, Mar 2014.

\bibitem{Jechow:11}
A.~Jechow, E.~W. Streed, B.~G. Norton, M.~J. Petrasiunas, and D.~Kielpinski.
\newblock Wavelength-scale imaging of trapped ions using a phase fresnel lens.
\newblock {\em Opt. Lett.}, 36(8):1371--1373, Apr 2011.

\bibitem{Shu:11}
Gang Shu, Chen-Kuan Chou, Nathan Kurz, Matthew~R. Dietrich, and Boris~B.
  Blinov.
\newblock Efficient fluorescence collection and ion imaging with the ``tack''
  ion trap.
\newblock {\em J. Opt. Soc. Am. B}, 28(12):2865--2870, Dec 2011.

\end{thebibliography}

\end{document}